\documentclass[aps,prd,twocolumn,showpacs,superscriptaddress,nofootinbib]{revtex4}  
\usepackage{graphicx}  
\usepackage{dcolumn}   
\usepackage{bm}        
\usepackage{amssymb}   
\usepackage{setspace}
\usepackage{hyperref}
\usepackage{textcomp}
\usepackage{amsmath}
\usepackage{amsfonts}
\begin{document}

\title{Big-bounce cosmology from  quantum gravity:  the case of cyclical Bianchi I Universe }

\author{Riccardo Moriconi}
\email{moriconi@na.infn.it}
\affiliation{Dipartimento di  Fisica "E. Pancini", Universit\`{a} di Napoli "Federico II",
 Compl. Univ. di Monte S. Angelo, Edificio G, Via Cinthia, I-80126, Napoli, Italy,}
\affiliation{Istituto Nazionale di  Fisica Nucleare (INFN) Sez. di Napoli, Compl. Univ. di Monte S. Angelo,
Edificio G, Via Cinthia, I-80126, Napoli, Italy,}

\author{Giovanni Montani}
\email{giovanni.montani@frascati.enea.it}
\affiliation{Dipartimento di Fisica (VEF), P.le A. Moro 5 (00185) Roma, Italy,}
\affiliation{ENEA, Unità Tecnica Fusione, ENEA C. R. Frascati, via E. Fermi 45, 00044 Frascati (Roma), Italy,}

\author{Salvatore Capozziello}
\email{capozziello@na.infn.it}
\affiliation{Dipartimento di  Fisica "E. Pancini", Universit\`{a} di Napoli "Federico II",
 Compl. Univ. di Monte S. Angelo, Edificio G, Via Cinthia, I-80126, Napoli, Italy,}
\affiliation{Istituto Nazionale di  Fisica Nucleare (INFN) Sez. di Napoli, Compl. Univ. di Monte S. Angelo,
Edificio G, Via Cinthia, I-80126, Napoli, Italy,} 
\affiliation{Gran
Sasso Science Institute (INFN),  Viale F. Crispi, 7, I-67100,
L'Aquila, Italy.}

\date{\today}

\begin{abstract}
We analyse the classical and quantum dynamics of a Bianchi I model in the
presence of a small negative cosmological constant  characterizing its
evolution in term of the dust-time 
dualism. 
We demonstrate that in a  canonical metric approach, the cosmological 
singularity is removed in correspondence 
to a positive defined value of the 
dust energy density. Furthermore, 
the quantum Big-Bounce is connected 
to the Universe turning point via 
a well-defined semiclassical limit. 
Then we can reliably infer that the 
proposed scenario is compatible with 
a cyclical Universe picture. 
We also show how, when the 
contribution of the dust energy density 
is sufficiently high, the proposed 
scenario can be extended to the Bianchi IX cosmology and therefore how it can 
be regarded as a paradigm for the 
generic cosmological model. 
Finally, we investigate the origin 
of the observed cut-off on the 
cosmological dynamics, demonstrating 
how the Big-Bounce evolution can be 
mimicked by the same semiclassical 
scenario, where the negative cosmological constant is replaced via a polymer 
discretization of the Universe volume. 
A direct proportionality law between 
such two parameters is then established. 
\end{abstract}

\maketitle
\section*{INTRODUCTION}

The Wheeler-DeWitt approach\cite{dewitt1},\cite{dewitt2},\cite{dewitt3}  to quantum cosmology \cite{misner},\cite{primordial} has two main relevant shortcomings, i.e. the absence of 
a unique definition of time \cite{isham}
and the difficulty in  
removing or properly interpreting the primordial singularity \cite{isham2},\cite{HH},\cite{sergei}.

Such problem, mainly characterizing all the canonical 
metric approaches, is essentially addressed by the Loop Quantum
Cosmology \cite{loop1},\cite{loop2},\cite{loop3}, where, adopting a scalar field as a relational time, it is shown the existence of a big bounce that remove the singularity.

However, this important result does not overcome some 
subtleties concerning its derivation and which are 
relevant on a general ground too. First of all, it is 
not clear if the choice of any relation time and, 
in particular the scalar field one, is suitable 
to describe the early Universe quantum dynamics\cite{zonetti},\cite{thiemann}.
Then it calls for attention the question concerning weather or not  
the symmetry preservation, characterizing Loop Quantum 
Cosmology, is the correct quantization procedure 
of a cosmological model \cite{LQCcianfrani}. 

The present paper analyses a cosmological model that 
contains features interesting for the deep understanding of the 
two points mentioned above. In fact, we consider a canonical 
minisuperspace model using a dust 
fluid as external time, according to the time-dust dualism discussed in \cite{KT1}. 
The very important feature of the obtained quantum cosmology 
is the emergence of a non-singular cyclical Universe, 
which is characterized by a quantum Big-Bounce and 
a classical turning point, associated to the existence 
of a small negative cosmological constant, i.e. 
small enough to ensure that such a re-collapsing feature 
be in the far future of the actual Universe. 

An important aspect of such a cosmological scenario, which legitimate the idea of cyclical Universe is the possibility 
to link the quantum evolution to the standard 
isotropic behaviour via a well-defined classical limit (see also \cite{odintsov},\cite{rev},\cite{mauro}, \cite{noether} for this problem in alternative theories of gravity). 
In fact the presence of a negative cosmological constant 
induces an harmonic oscillator morphology to the 
system Hamiltonian (a part a global minus sign) and 
this is responsible both for the existence of a classical 
limit and of the positive nature of the dust energy density.
This latter fact solves, in our cosmological implementation, 
the basic problem of the approach discussed in \cite{KT1}.

More in detail, we consider the evolutionary quantum dynamics 
of a Bianchi I model in the presence of a negative 
cosmological constant, as represented in Misner-like  variables \cite{misnermixmaster},\cite{gravitation}. 
Clearly, the classical limit corresponds to an increasingly 
isotropic Universe, although we do not address here the 
role of the matter and then the reproduction of 
Standard Cosmology. This is because, we aim to determine 
a cosmological behaviour which be able to mimic a very 
general cosmological scenario near the singularity, 
according to the idea that the natural isotropization mechanism 
must be recogniced in the inflationary scenario\cite{montani-kirillov}.

To this end, we investigate the implications of our 
dynamical model on the evolution of the Bianchi IX 
cosmology, which is, accordingly to the Belinski-Khalatnikov-Lifshitz (BKL) conjecture, the prototype for the evolution of a generic inhomogeneous Universe on a sufficiently small spatial scale \cite{BKL1982}. We demonstrate that, along the dynamics of the stable expectation values of the configurational variables, the presence of the Bianchi IX potential can be neglected, as soon as the value of the dust energy density is sufficiently large. Thus, for such a (non-severe) restriction, the Bianchi I and Bianchi IX model quantum dynamics overlap nearby the primordial singularity and our result acquires a high degree of generality, i.e. 
our picture of a cyclical Universe could have 
a very general implementation in the generic 
cosmological problem. Finally, we investigate which ingredient of our model is relevant in determining a cut-off physics and we show that there exists a direct relation between the negative cosmological constant presence and an effective semiclassical polymer dynamics \cite{corichi},\cite{corichidue}, in which that constant is removed but the discrete nature of the Universe volume is included. 

Summarizing, the present paper discuss a cosmological 
scenario containing a number of very peculiar 
properties, suggesting that its features are 
physically meaningful and are not 
formal coincidences. In particular, we stress how, 
in the present canonical evolutionary quantum context, 
the emergence of a Big-Bounce and of a cyclical Universe 
is at all natural and general in its structure, 
so much to encourage more general implementations.

This paper is organized as follows.

In Section \ref{sec:BI} we describe the Bianchi I model in presence of a negative cosmological constant from the classical and from the quantum point of view. The first part of the Section is devoted to analyse the classical trajectories of the Misner-like variables near the singularities while in the second part we compare this classical behaviours with the related quantum expectation values.

In Section \ref{sec:BIX} we generalize, in a qualitatively way, the properties founded for the Bianchi I model to the more general Bianchi IX model, shedding light on the role playing by the potential term.

The Section \ref{sec:phen} is dedicated to the cosmological interpretation of the results obtained in the previous, giving in particular a phenomenological explanation of how to extend the features of the Bianchi I and Bianchi IX model to the generic inhomogeneous Universe.

Then, in Section \ref{sec:poly}, we see how the role of the negative cosmological constant is related to a cut-off physics, making use of a polymer quantization for the variable connected to the Universe volume.

Brief concluding remarks complete the paper.

\section{Bianchi I quantum dynamics in the Kucha\v{r} and Torre Approach}
\label{sec:BI}

The cosmological scenario we are going to implement can 
be applied also to the isotropic Universe \cite{pittorino}, 
as soon as the role played here by the anisotropy variables
is supplied by a massless (or even self-consistent) scalar 
field. Indeed, the kinetic term in the Hamiltonian of a 
scalar field on the isotropic Universe dynamics is 
at all isomorphic to that one of an anisotropic variable 
in the Misner representation 
(\textit{i.e.} $\beta _+$ or $\beta _-$) in the Hamiltonian of 
a Bianchi Model, in particular for the type I and IX we 
will address in this paper. 
The motivation to consider the present more general scheme 
than the isotropic Universe must be individualized in the 
natural presence of the anisotropy terms near the cosmological 
singularity, in comparison to the necessity of postulating 
the presence of a kinetic scalar field contribution asymptotically to the singularity (a reasonable but not rigorously proved 
feature associated to the pre-inflationary inflaton dynamics 
\cite{primordial}). 
Furthermore, the morphology of the Bianchi I and IX 
models outlines a high degree of generality with respect to 
the Robertson-Walker geometry since, as shown in \cite{BKL1982}, 
the generic cosmological solution, near the singularity, is 
an infinite series of Kasner epochs (periods of time 
in which the dynamics is Bianchi I-like), one for each 
space point (physically for each cosmological horizon). 
Such a basic result, known as the BKL conjecture, suggests that the analysis here addressed can 
be implemented to a very general picture and we can infer that 
the discussed scenario removes the cosmological singularity 
for a generic inhomogeneous Universe, as far as its evolution 
admits the Bianchi IX oscillatory regime as a homogeneous prototype. 
In what follows, we prefer to deal with minisuperspace models, 
in order to avoid the non-trivial question of 
how can be rigorously implemented the conjecture above on 
a quantum level: the spatial decoupling of the space 
point in the asymptotic dynamics of an inhomogeneous Universe 
towards the singularity is demonstrated in the classical 
sector, on the base of statistical arguments \cite{kirillovsta}, 
but it remains an open issue in a metric quantum dynamics. 
Let us consider a Universe described by a Bianchi I model in the presence of a negative cosmological constant $-\Lambda$, with $\Lambda>0$. It is useful to describe the model with respect to the Misner variables $\{\alpha,\beta_{\pm}\}$, where $\alpha$ expresses the isotropic volume of the universe (the initial singularity is reached for $\alpha\rightarrow -\infty$) while $\beta_{\pm}$ accounts for the anisotropies of this model. In the Appendix (\ref{appendix}) we provide a brief derivation to show that the associated minisuperspace superHamiltonian takes the form\footnote{We use the $(-,+,+,+)$ signature of the metric and the geometric unit system $(c=G=\hbar=1)$.}
\begin{equation}
\label{hamiltoniana}
\mathcal{H} = \frac{e^{-3\alpha}}{24\pi}\left[ -p_{\alpha}^{2} +p_{+}^{2} +p_{-}^{2} \right] -\pi e^{3\alpha}\Lambda,
\end{equation}
where $ \{ p_{\alpha},p_{+},p_{-} \} $ are the conjugated momenta related to the Misner variables.
In view of a later quantization of the model, it is convenient to introduce the auxiliary variable $\rho$ such that:
\begin{equation}
\rho = e^{\frac{3}{2}\alpha} \quad \longrightarrow \quad p_{\rho} = \frac{2}{3}e^{-\frac{3}{2}\alpha}p_{\alpha}.
\end{equation}
In terms of this new conjugated variables the superHamiltonian (\ref{hamiltoniana}) takes the form
\begin{equation}
\label{hamrho}
\mathcal{H} = -\frac{3}{32 \pi} p_{\rho}^{2} + \frac{p_{+}^{2} +p_{-}^{2}}{24 \pi \rho ^{2}} -\pi \rho ^{2} \Lambda.
\end{equation}
We now perform a canonical quantization of the system, after the definition of a suitable Hilbert space, by replacing the space-phase variables with multiplicative operators for variables $\{ \rho,\beta_{+},\beta_{-}\}$ and differential operators for $ \{ p_{\rho},p_{+},p_{-}\}$, so that:
\begin{equation}
p_{i}\rightarrow\ -i \frac{d}{dq_{i}} \quad , \quad q_{i} = \{  \rho,\beta_{+},\beta_{-}\}.
\end{equation}
If now we introduce the wave function of the Universe $\psi( \rho,\beta_{\pm})$ we can apply to it the quantum version of the superHamiltonian (\ref{hamrho}) in order to obtain the Wheeler-deWitt operator
\begin{equation}
\label{WdW}
\hat{\mathcal{H}}\psi( \rho,\beta_{\pm}) = \left[\frac{3}{32 \pi} \partial_{\rho}^{2} - \frac{\partial_{+}^{2}+\partial_{-}^{2}}{24 \pi \rho ^{2}} -\pi \rho ^{2} \Lambda \right]\psi( \rho,\beta_{\pm}).
\end{equation}
\subsection{Evolutionary quantum cosmology}
Here we take into account the evolutionary quantum theory, as it is analysed in \cite{KT1},\cite{KT2}. In these works it is considered a system of normal Gaussian coordinates $X^{\mu} = (T,X^{i})$, or in other words a synchronous reference system, for which the line element of the metric takes the form:
\begin{equation}
\label{elementolineasinchro}
ds^{2}=-dT^{2}+h_{ij}dX^{i}dX^{j}, 
\end{equation}
where the indices $\{ i,j\}$ are summed over the spatial directions and $h_{ij}$ is the spatial metric. In this way four components of  the space-time metric $g_{\mu \nu}$ are fixed by the \textit{Gaussian conditions}:
\begin{equation}
\label{temp}
g_{00} + 1 =0 \quad , \quad g_{0i}=0.
\end{equation}
The physical meaning of the previous conditions is more clear in the context of the ADM (Arnowitt-Deser-Misner \cite{ADM}) formalism, for which the space-time metric $g_{\mu\nu}$ is replaced by the \textit{lapse function} $N$, the \textit{shift vector} $N^{i}$ and the spatial metric $h_{ij}$.
In the ADM procedure we perform a foliation of the space-time: the lapse function $N$ represents the proper time separation between two neighboring leaves, while the shift vector $N^{i}$ represents the displacement, with respect to a normal projection, of the local spatial coordinate system in the intersection with the successive leave. In the ADM formalism the space-time metric takes the form:
\begin{equation}
\label{elemtnolineaADM}
ds^{2} = N^{2}dt^{2} - h_{ij}(N^{i}+dx^{i})(N^{j}+dx^{j}).
\end{equation}
If we make a comparison between the line elements (\ref{elementolineasinchro}) and (\ref{elemtnolineaADM}) it is clear that the conditions (\ref{temp}) are equivalent to
\begin{equation}
\label{ADMcondition}
N = 1 \quad , \quad N^{i}=0,
\end{equation}
where the foliation of the space-time is such that $t=T$ and $x^{i}=X^{i}$.
The relations (\ref{ADMcondition}) tell us that everywhere the proper time between two neighboring leaves is the same and that there is no displacement, with respect to the normal projection, between one leaves and another. 
If now we want to implement the Gaussian conditions in the action principles of general relativity, for example in the vacuum case, we can follow two ways: in the first one we impose the conditions after the variation of the Einstein-Hilbert action, while in the other case we adjoin them to the action, making use of Lagrangian multipliers technique, before the variation. 

When we proceed in the first manner, we deal with the Einstein-Hilbert Action in vacuum 
\begin{equation}
\label{azionebase}
S^{G} = -\frac{1}{2\kappa}\int d^{4}x\sqrt{-g}R,
\end{equation}
and a variation of this action with respect to the space-time metric $g_{\mu \nu}$ leads to the Einstein equations in vacuum:
\begin{equation}
G_{\mu \nu} = R_{\mu \nu}- \frac{1}{2}g_{\mu \nu}R = 0.
\end{equation}
An equivalent form of the action (\ref{azionebase}) is obtained in the ADM formalism, for which we have
\begin{equation}
S^{G}[h_{ij},N,N^{i}] = \int _{\mathbb{R}}dt \int _{\Sigma}d^{3}x \left[ \dot{h}_{ij}P ^{ij} - ( N^{i}\mathcal{H}_{i}^{G} + N\mathcal{H} ^{G}) \right]
\end{equation}
where
\begin{equation}
\label{superhamil}
\mathcal{H}^{G} = \mathcal{G}_{ijkl}P^{ij}P^{kl} - \frac{\sqrt{h}}{2k}\overline{R}
\end{equation}
\begin{equation}
\label{supermomento}
\mathcal{H}_{i}^{G} = -2h_{ik}\nabla _{j}P^{kj}
\end{equation}
\begin{equation}
\label{supermetrica}
\mathcal{G}_{ijkl} = \frac{k}{\sqrt{h}}(h_{ik}h_{jl} + h_{jk}h_{il} - h_{ij}h_{kl})
\end{equation}
are respectively the \textit{superHamiltonian}, the \textit{supermomentum}, the \textit{supermetric} and $P^{ij}$ is the conjugated momenta to the spatial metric $h_{ij}$.
The variation with respect to $N$ and $N^{i}$ gives the secondary constraints:
\begin{equation}
\label{vincoli}
\mathcal{H}^{G} = \mathcal{H}_{i}^{G} = 0.
\end{equation}
The Hamilton equations for $h_{ij}$ and $P_{ij}$, once fixed $N=1$ and $N^{i}=0$, provide, together with the constraints (\ref{vincoli}), the Einstein equations in the synchronous reference frame. 

The second way to proceed consists of adding the coordinate conditions (\ref{temp}) in the Einstein-Hilbert action by the multipliers $M$ and $M_{i}$ in such a way that an extra term $S^{F}$ appears in the action:
\begin{equation}
\label{azionemolt}
S[g_{\mu \nu},M,M_{k}] = S^{G} + S^{F},
\end{equation}
with
\begin{multline}
S^{F}[g_{\mu \nu},M,M_{k}]= \\ -\frac{1}{2\kappa}\int  d^{4}x \left[ -\frac{1}{2}M\sqrt{-g}(g^{00} + 1 ) + M_{i}\sqrt{-g}g^{0i}  \right]
\end{multline}
and where we defined the quantity:
\begin{equation}
\begin{cases}
\label{moltiplicatorilagrange}
& M:=-\frac{H^{G} }{\sqrt{h}},\\
& M_{i}:= \frac{H_{i}^{G}}{\sqrt{h}}.
\end{cases}
\end{equation}
Clearly the variation of the action (\ref{azionemolt}) introduces a source term in the Einstein equations. 
The role of Lagrangian multipliers $M$, $M_{k}$ is clear if we write the action (\ref{azionemolt}) in the ADM formalism, in order to obtain:
\begin{multline}
\label{azionemoltADM}
S[h_{ab},N,N^{i},M,M_{k}] = \\ = \int _{\mathbb{R}}dt \int _{\Sigma}d^{3}x [ \dot{h}_{ij}P ^{ij} - ( N^{i}\mathcal{H}_{i}^{G} + N\mathcal{H} ^{G}) + \\ -\frac{1}{2}M\sqrt{h}(N- N^{-1}) + M_{i}\sqrt{h}NN^{i}].
\end{multline}
If we perform a variation by  $M$ and $M_{i}$ we obtain the Gaussian conditions (\ref{ADMcondition}), while a variation with respect to $N$ and $N^{i}$ gives the Eq.'s (\ref{moltiplicatorilagrange}) and fix the multipliers $M$ and $M_{i}$ as functions of the canonical variables $h_{ij},P^{ij}$. If we use the Eq's (\ref{ADMcondition}) and (\ref{moltiplicatorilagrange}) to eliminate the presence of the mutipliers $N$, $N^{i}$ and $M$, $M_{i}$, the action (\ref{azionemoltADM}) clearly reduces to the canonical action (\ref{azionemolt}).

Looking at the action (\ref{azionemolt}), it is not invariant under arbitrary transformations of space-time coordinates and this is due to the fact that we have introduced a privileged coordinate system, \textit{i.e} the normal Gaussian coordinates. However, it is always possible to restore the diffeomorphism invariance making a parametrization of the coordinates. It means that if we take the Gaussian coordinates as a functions of a arbitrary coordinates $x^{\alpha}$ in such a way that $X^{\mu} = (T(x^{\alpha}),X^{i}(x^{\alpha}))$ the action (\ref{azionemolt}) can be expressed as:
\begin{multline}
\label{azionemoltpar}
S[g_{\alpha \beta},M,M_{k},X^{\mu}] = S^{G} + S^{F} = -\frac{1}{2\kappa}\int d^{4}x\sqrt{-g}R  + \\ -\frac{1}{2\kappa}\int  d^{4}x \sqrt{-g}\left[ -\frac{1}{2}M(g^{\alpha \beta}T_{,\alpha}T_{,\beta} + 1 ) + M_{i}g^{\alpha \beta}T_{,\alpha}X_{,\beta}^{i}  \right],
\end{multline}
that is manifestly invariant under arbitrary transformations of $x^{\alpha}$.

The form of the action (\ref{azionemoltpar}) allows us to understand the nature of the source of the gravitational field, described by that part of the action appearing in the second row. In \cite{KT1} this source term is defined as \textit{Gaussian Reference Fluid}.

The variation of the action (\ref{azionemoltpar}) by the metric $g_{\alpha\beta}$ gives the Einstein equations:
\begin{equation}
G_{\alpha \beta} = \kappa T_{\alpha \beta},
\end{equation}
where 
\begin{equation}
T^{\alpha \beta}=\frac{2}{\sqrt{-g}}\frac{\delta S^{F}}{\delta g_{\alpha \beta}}
\end{equation}
is the energy-momentum tensor associated to the reference fluid.
After the definition of the four-velocity vector
\begin{equation}
\label{4-velocità}
U^{\alpha}:=-g^{\alpha \beta}T_{,\beta},
\end{equation}
it is possible to evaluate the energy-momentum tensor in order to give a clear physical interpretation of the presence model:
\begin{equation}
\label{TEI}
T^{\alpha \beta} =M U^{\alpha}U^{\beta} + M^{(\alpha}U^{\beta)}.
\end{equation}
The Eq.(\ref{TEI}) is equivalent to the Eckart energy-momentum tensor\cite{eckart} that describes a heat-conducting fluid. The absence of a stress part in the energy-momentum tensor tells us that the Gaussian reference fluid behaves as a dust. In particular, if we impose only the time condition ($M^{i}=0$) the Eq.(\ref{TEI}) becomes:
\begin{equation}
\label{TEIdust}
T^{\alpha \beta} =M U^{\alpha}U^{\beta},
\end{equation}
which describes the behaviour of an incoherent dust, where $M$ is the rest mass density and $U^{\alpha}$ is the four-velocity. 

If now we consider the canonical ADM form of the action (\ref{azionemoltpar}) we have 
\begin{multline}
\label{azioneADMdust}
S[h_{ij},X^{\mu},M,M_{k}] = \int _{\mathbb{R}}dt \int _{\Sigma}d^{3}x [ \dot{h}_{ij}P ^{ij} + \dot{X}^{\mu}P_{\mu} + \\ - ( N^{i}\mathcal{H}_{i} + N\mathcal{H}) ,
\end{multline}
with
\begin{equation}
\mathcal{H} = \mathcal{H}^{G} +\mathcal{H}^{D} \quad , \quad \mathcal{H}_{i} = \mathcal{H}_{i}^{G} + \mathcal{H}_{i}^{D}.
\end{equation}
where $P_{\mu} = (P,P_{i})$ are the conjugated momentas to $X_{\mu} = (T,X_{i})$. The quantity $\mathcal{H}^{D}$ and $\mathcal{H}^{D}_{i}$ are respectively the superHamiltonian and supermomentum contribution due to the reference fluid and, when we take into account the case of an incoherent dust, they simply becomes:
\begin{equation}
\mathcal{H}^{D} = P \quad , \quad \mathcal{H}^{D}_{i}= X^{j}_{,i}P_{j}=0.
\end{equation}
As before, the variation with respect to $N$ and $N^{i}$ gives us the constraints:
\begin{equation}
\label{vincoloSHdust}
\mathcal{H} = \mathcal{H}^{G} + \mathcal{H}^{D} = \mathcal{H}^{G} + P = 0, 
\end{equation}
\begin{equation}
\label{vincoloSMdust}
\mathcal{H}_{i} = \mathcal{H}^{G}_{i} + \mathcal{H}^{D}_{i} = \mathcal{H}^{G}_{i} = 0.
\end{equation}

The quantization procedure of the system composed by an incoherent dust coupled with gravity\cite{KT1} consists to associate to the canonical variables the following differential operators
\begin{equation}
\label{operatorispa}
\hat{h}_{ij} = h_{ij}\times \quad , \quad \hat{P}^{ij} = -i\frac{\delta}{\delta h_{ij}},
\end{equation}
\begin{equation}
\label{operatoridust}
\hat{X}^{\mu} = X^{\mu}\times \quad , \quad \hat{P}_{\mu} = -i\frac{\delta}{\delta X^{\mu}},
\end{equation}
and to evaluate the action of the quantum version of the constraints (\ref{vincoloSHdust}),(\ref{vincoloSMdust}) on the physical states identified as the functional $\Psi[X^{\mu},h_{ij}]$, \textit{i.e.} the wave function of the Universe.

First of all, the condition $\mathcal{H}^{D}_{i}= X^{j}_{,i}P_{j}=0$ tells us that
\begin{equation}
\frac{\delta}{\delta X^{i}}\Psi[X^{\mu},h_{ij}] = 0,
\end{equation}
so the wave function of the Universe does not depend on the spatial fluid variables $X^{i}$ but only on the time fluid variable $T$.
Furthermore, the quantum version of the constraint (\ref{vincoloSMdust}),
\begin{equation}
\hat{\mathcal{H}}_{i}\Psi[T,h_{ij}] = 0,
\end{equation}
ensures us that $\Psi[T,h_{ij}]$ does not depend on the particular metric representation, but only on 3-geometries.

Remembering the definitions of the operators (\ref{operatorispa}),(\ref{operatoridust}), the application of the constraint (\ref{vincoloSHdust}) on the physical states $\Psi[T,h_{ij}]$ leads us to the Wheeler-DeWitt(WDW) equation that resembles a Schrodinger-like equation:
\begin{multline}
\label{WDWdust}
\hat{\mathcal{H}}\Psi[T,h_{ij}] = \left[\hat{\mathcal{H}}^{G} - i\frac{\delta}{\delta T}\right]\Psi[T,h_{ij}] = 0 \rightarrow \\ \rightarrow  i\frac{\delta}{\delta T}\Psi[T,h_{ij}] = \hat{\mathcal{H}}^{G}\Psi[T,h_{ij}],
\end{multline}
which determines the evolution of the system with respect to the time variable $T$.
It is easy to verify that a general solution for the Eq.(\ref{WDWdust}) is
\begin{equation}
\Psi(T,h_{ij}) = \int dE \psi(E,h_{ij}) e^{-i E T},
\end{equation}
leading to the time independent eigenvalue problem
\begin{equation}
\label{problemaautovalori}
\hat{\mathcal{H}}^{G}\psi = E \psi.
\end{equation}
From the Eq.(\ref{problemaautovalori}) we can see that $E$ is the eigenvalue of the superHamiltonian, and it is associated to the dust energy density via the relation $\rho_{dust} = - \frac{E}{\sqrt{h}}$. 
For the Bianchi I model that we are taking into account the superHamiltonian $\mathcal{H}^{G}$ is of the form (\ref{hamrho}), which in the quantum version $\hat{\mathcal{H}}^{G}$ correspond to the Eq.(\ref{WdW}), and the eigenvalue problem (\ref{problemaautovalori}) takes the explicit form:
\begin{equation}
\label{WdWe}
\left[\frac{3}{32 \pi} \partial_{\rho}^{2} - \frac{\partial_{+}^{2}+\partial_{-}^{2}}{24 \pi \rho ^{2}} -\pi \rho ^{2} \Lambda \right]\psi( \rho,\beta_{\pm}) = E \psi( \rho,\beta_{\pm}).
\end{equation}

The Kucha\v{r} and Torre approach is clearly a promising 
point of view for addressing the problem of time, 
viewed as a necessary weakening of the General Relativity 
Principle. Indeed, although the general covariance is 
preserved via a general reparametrization, the time 
evolution of the quantum gravitational field comes 
out from the privileged character of the Gaussian reference 
frame. But the real critical point of the formulation presented above is that the super-Hamiltonian spectrum is not positive 
defined and consequently the dust fluid has to 
possess a non-positive energy density: a really unpleasent 
physical property, which is a serious shortcoming of the 
formulation. In \cite{KT2}, it has been demonstrated 
that a real inchorent dust coupled to gravity play 
the role of a physical clock and this issue constitutes 
a complementary approach to the present one. 

A part from the non-trivial question about how it is 
possible to make the Gaussian frame compatible with the 
energy conditions \cite{KT1} (\textit{i.e.} its energy momentum 
tensor does not fulfill the condition to represent a 
physical fluid), we can see that a dualism exists 
between a physical clock for the gravitational field 
and a fluid of reference coupled to the gravitational 
field dynamics, see also \cite{Montani2002},\cite{mercuri},\cite{mercuri2004}. From a more general point of view, we can  
infer that the coupling of the gravitational field to 
a given physical fluid is equivalent to induce 
a no longer vanishing super-Hamiltonian and/or super-momentum 
constraints. 
From a field theory point of view, we are arguing that 
the quantization of the gravitational field is affected by 
the choice of a specific gauge, \textit{i.e.} of a real system of 
reference, by restoring a time evolution. 
In quantum gravity, the distinction between a 
real reference frame (a physical system) having 
a non-zero energy-momentum tensor, and a simple system 
of coordinates (a mathematical reparametrization of 
the dynamics) is deep: while in General Relativity the 
two concepts overlap, as soon as, we take the real 
fluid as a test system, on the quantum level, the 
energy-momentum tensor of the reference frame participate 
the gravitational field dynamics via the super-Hamiltonian 
spectrum. 

The present study addresses the question concerning the 
positive character of the dust energy density, since 
we construct a quantum cosmology model for which such property definitely holds.  It is actually relevant that from such a regularization of the Kukha\v{r} and Torre model the relevant issues described below come out: the emergence of a cyclical Universe, 
possessing a Big-Bounce feature and the proper classical limit.
The basic ingredient for such a physical implementation 
of the clock-dust dualism is the presence of a small 
negative cosmological constant (also ensuring the Universe 
turning point), while the evolutionary quantum dynamics 
is then crucial for the emergence of a cyclical picture. 
The physical meaning of our cosmological time consists 
of the possibility to restate the Bianchi I super-Hamiltonian 
eigenvalue as the energy density of a physical fluid, 
comoving with the syncrhronous reference system and, de facto, 
identified with the latter. In the classical limit, 
our Universe possesses a dust contribution (non-relativistic 
matter) which is redshifted by the inflationary paradigm 
\cite{primordial},\cite{kolb} up to so much small 
values that its present day contribution to the Universe 
critical parameter is at all negligible, see 
\cite{pittorino},\cite{Battisti2006},\cite{Battisti2007},\cite{Battisti2008}.
By other words, our physical dust is a valuable clock 
to describe the considered model evolution, but it is today 
so much diluted across the Universe that the difference with 
a formal system of coordinates is no longer appreciable 
and the General Relativity Principle is fully restored.  

\subsection{Semiclassical limit}
Before to deal with the full quantum problem, it is interesting for our porpoises to study the associated classical problem to the Eq.(\ref{problemaautovalori}), namely the zero-th order of a WKB expansion of the evolutionary quantum system\cite{WKBmontani}. 
The constraint that we obtain is
\begin{equation}
\label{sH}
\mathcal{H}=-\frac{3}{32 \pi} p_{\rho}^{2} + \frac{p_{+}^{2} +p_{-}^{2}}{24 \pi \rho ^{2}} -\pi \rho ^{2} \Lambda = E
\end{equation}
We can solve the classical dynamics making use of the Hamiltonian equations and the constraint ($\ref{sH}$). We can find solution for the isotropic variable $\rho$ taking into account the Hamiltonian equations\footnote{In the following we label the Gaussian time variable $T$ as $t$.}
\begin{equation}
\label{eqhamiltonrho}
\begin{cases}& \dot{\rho}=\frac{d \rho}{d t}=\frac{\partial \mathcal{H}}{\partial p_{\rho}} = -\frac{3}{16\pi}p_{\rho} \\ & \dot{p_{\rho}}=\frac{d p_{\rho}}{d t}=-\frac{\partial \mathcal{H}}{\partial \rho} = \frac{p_{+}^{2} +p_{-}^{2}}{12 \pi \rho ^{3}} + 2\pi \rho \Lambda, \end{cases} 
\end{equation}
in order to obtain
\begin{equation}
\label{moto}
\ddot{\rho} + \frac{p_{+}^{2} +p_{-}^{2}} {64 \pi^{2} \rho ^{3}} + \frac{3}{8} \rho \Lambda = 0.
\end{equation} 
Recalling that $p_{\rho}=-\frac{16\pi}{3}\dot{\rho}$, the superHamiltonian constraint become
\begin{equation}
\label{vincolo}
\dot{\rho}^{2} - \frac{p_{+}^{2} +p_{-}^{2}}{64 \pi^{2} \rho ^{2}} +\frac{3}{8} \rho^{2} \Lambda + \frac{3}{8\pi}E=0.
\end{equation}
It is possible to show that a solution for Eq.'s (\ref{moto}) and (\ref{vincolo}) is given by
\begin{multline}
\label{eq.moto rho}
\rho(t) =\\ = \sqrt{\left(\frac{-E}{2\pi\Lambda}\right)\left[ 1 \pm \sqrt{1+\frac{\Lambda (p_{+}^{2} + p_{-}^{2})}{6E^{2}}}\sin
\left( \sqrt{\frac{3\Lambda}{2}}t + \varphi \right)\right]}.
\end{multline}
The solution (\ref{eq.moto rho}) exhibits the usual initial singularity in the past for which $\rho=0 \rightarrow \alpha =-\infty$ and furthermore a singularity in the future exists too, namely a \textit{big crunch} singularity. The value of the integration constant $\varphi$ can be chosen in such a way that the initial singularity happen for the value $t=0$, to give us:
\begin{equation}
\varphi_{0} = \arcsin \left(\mp\frac{1}{\sqrt{1+\frac{\Lambda (p_{+}^{2} + p_{-}^{2})}{6E^{2}}}}\right).
\end{equation}
\begin{figure}[h!]
\centering
\includegraphics[scale=.67]{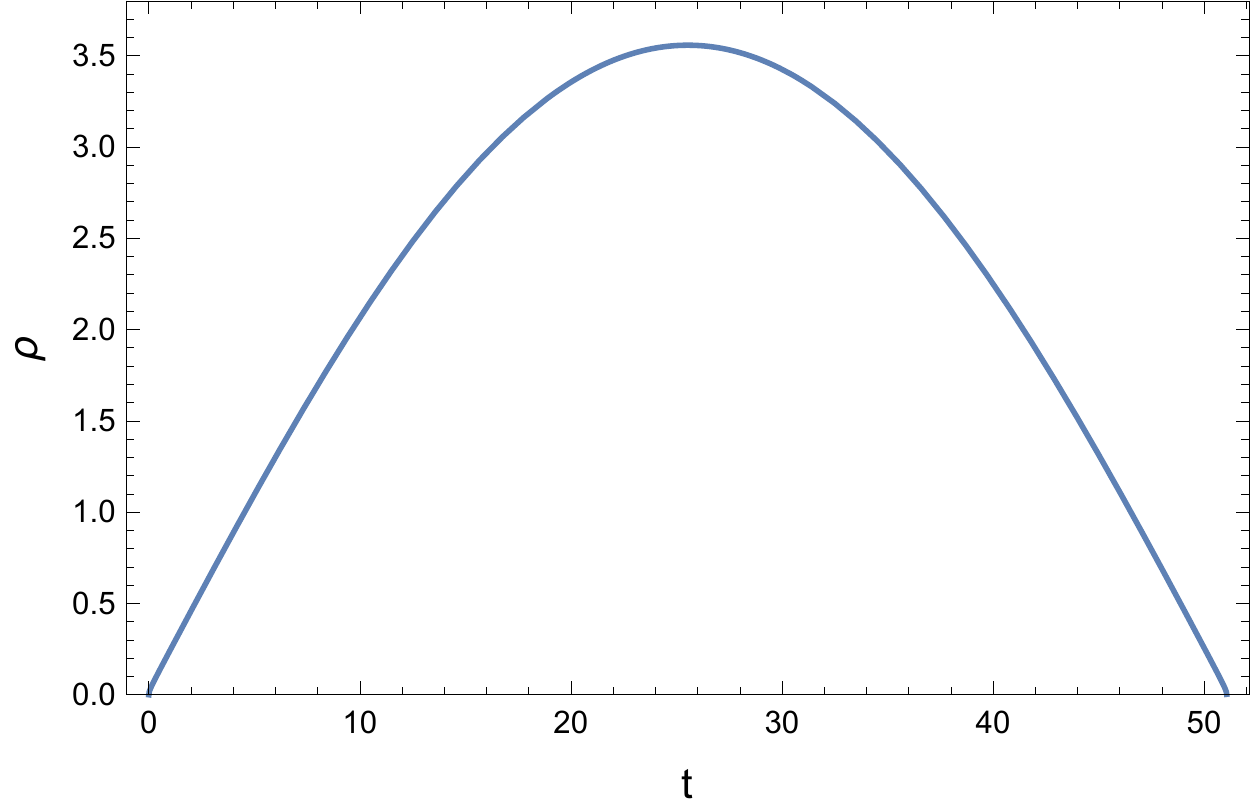}
\caption{\footnotesize The classical trajectory for the isotropic variable $\rho$ exhibit a singularity in the past and another one in the future. The solution is sketched for the parameters: $\Lambda =0.01, p_{+}=p_{-}=0.1, E=-0.397$.}
\label{fig:rho}
\end{figure}

\begin{figure}[h!]
\centering
\includegraphics[scale=.67]{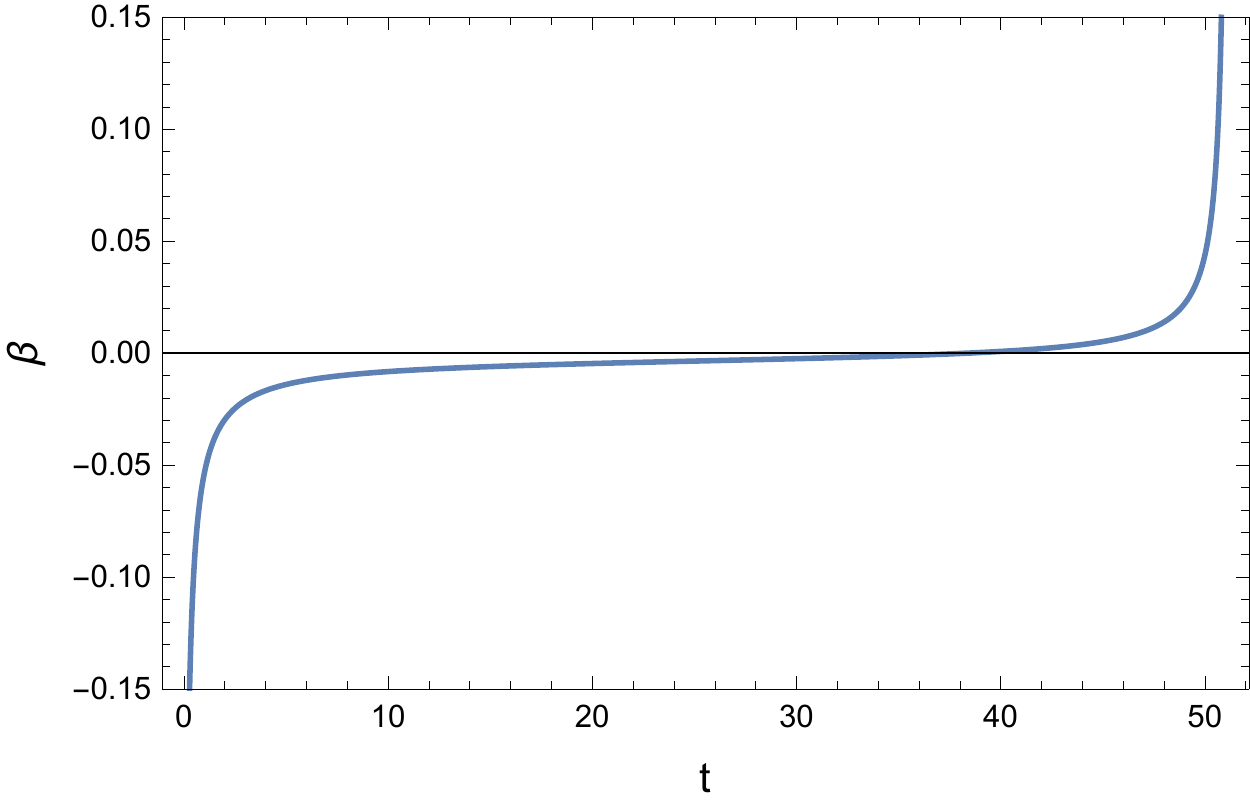}
\caption{\footnotesize The classical trajectory for the anisotropies $\beta_{\pm}$. Next to the singularities the anisotropies diverge. The solution is sketched for the parameters: $\Lambda =0.01, p_{+}=p_{-}=0.1, E=-0.397$.}
\label{fig:beta}
\end{figure}
The classical behaviour of the isotropic variable $\rho$ is sketched in Fig.(\ref{fig:rho}).
Analogously, The classical dynamics of the anisotropies $\beta_{\pm}$ can be solved, including the solution (\ref{eq.moto rho}) inside the hamiltonian equations. This way, we have
\begin{equation}
\label{eqhamiltonbeta}
\begin{cases}& \dot{\beta_{\pm}}=\frac{\partial \mathcal{H}}{\partial p_{\pm}} = \frac{p_{\pm}}{12\pi\rho^{2}} = \\ &  =-\frac{\Lambda p_{\pm}}{6E}\left[1 \pm \sqrt{1+\frac{\Lambda (p_{+}^{2} + p_{-}^{2})}{6E^{2}}}\sin
\left( \sqrt{\frac{3\Lambda}{2}}t + \varphi_{0} \right)\right]^{-1} \\ & \dot{p_{\pm}}=-\frac{\partial \mathcal{H}}{\partial \beta_{\pm}} = 0 \end{cases} 
\end{equation}
The solution reads as
\begin{widetext}
\begin{equation}
\label{eq.moto beta}
\beta_{\pm}(t) = \frac{p_{\pm}}{3\sqrt{p_{+}^2 + p_{-}^2}} \ln \left| \frac{1+ \frac{\sqrt{6}E}{\sqrt{\Lambda(p_{+}^2 + p_{-}^2)}}\left( \sqrt{1+\frac{\Lambda (p_{+}^{2} + p_{-}^{2})}{6E^{2}}} + \tan \left( \frac{1}{2}\sqrt{\frac{3\Lambda}{2}}t + \frac{\varphi_{0}}{2} \right) \right)}{ 1- \frac{\sqrt{6}E}{\sqrt{\Lambda(p_{+}^2 + p_{-}^2)}}\left( \sqrt{1+\frac{\Lambda (p_{+}^{2} + p_{-}^{2})}{6E^{2}}} + \tan \left( \frac{1}{2}\sqrt{\frac{3\Lambda}{2}}t + \frac{\varphi_{0}}{2} \right) \right) } \right| + const. .
\end{equation}
\end{widetext}
As we can see in  Fig.(\ref{fig:beta}), at the classical level the anisotropies of the model become important in magnitude towards the singularities in the past and in the future. So, the presence of a negative cosmological constant in the semiclassical evolution case do not mine the divergence of the anisotropies towards the singularities, typical of the anisotropic models.
\subsection{Dynamics of the quantum expectation values}
\label{sec:valorimedi}
\begin{figure}[h!]
\centering
\includegraphics[scale=.67]{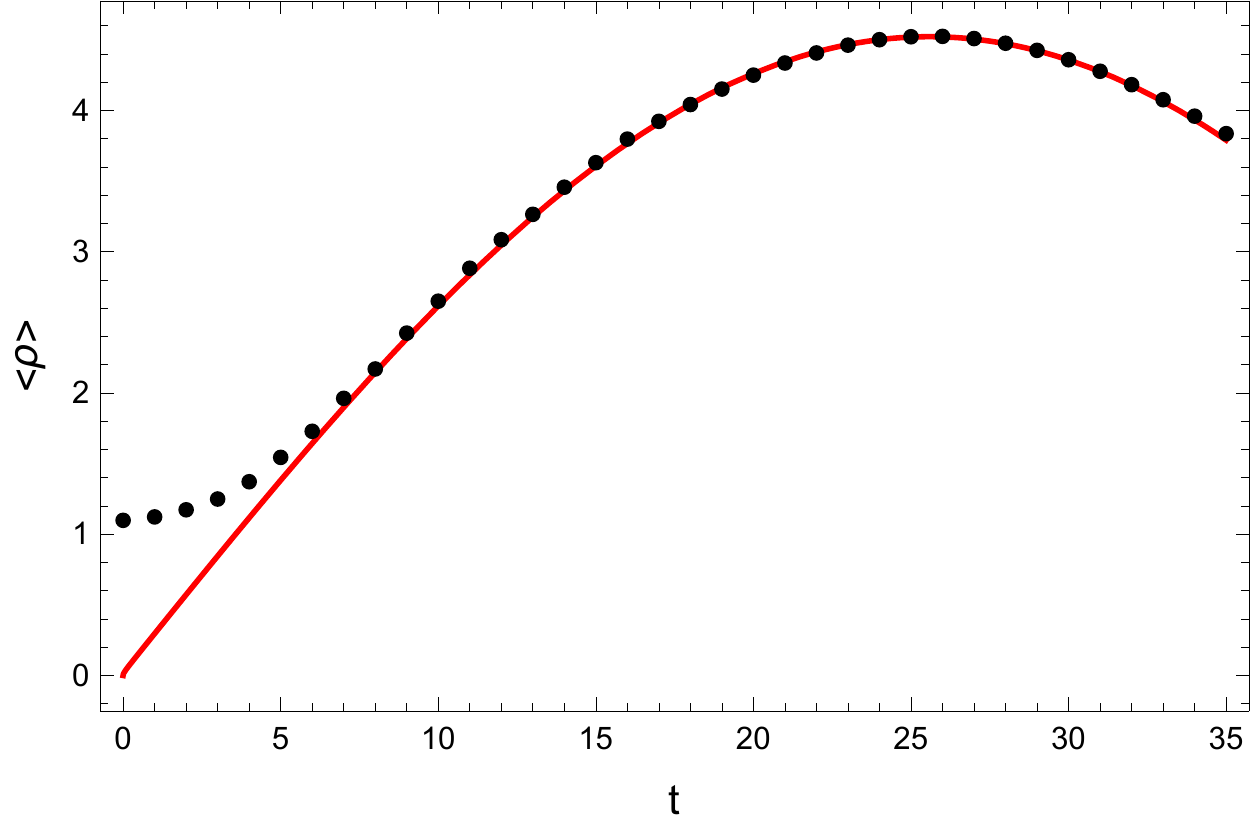}
\caption{\footnotesize The black points represent the expectation value $<\rho>_{t}$ evaluated via numerical integration for the following choose of the integration parameters: $\Lambda =0.01, k^{'*}=5, k^{*}_{+}=k^{*}_{-}=0.1, \sigma_{+}=\sigma_{-}=0.01, \sigma=0.88$. The continuous red line represents the classical trajectory evaluated with the same classical parameters.}
\label{fig:pacchettirho}
\end{figure}

\begin{figure}[h!]
\centering
\includegraphics[scale=.67]{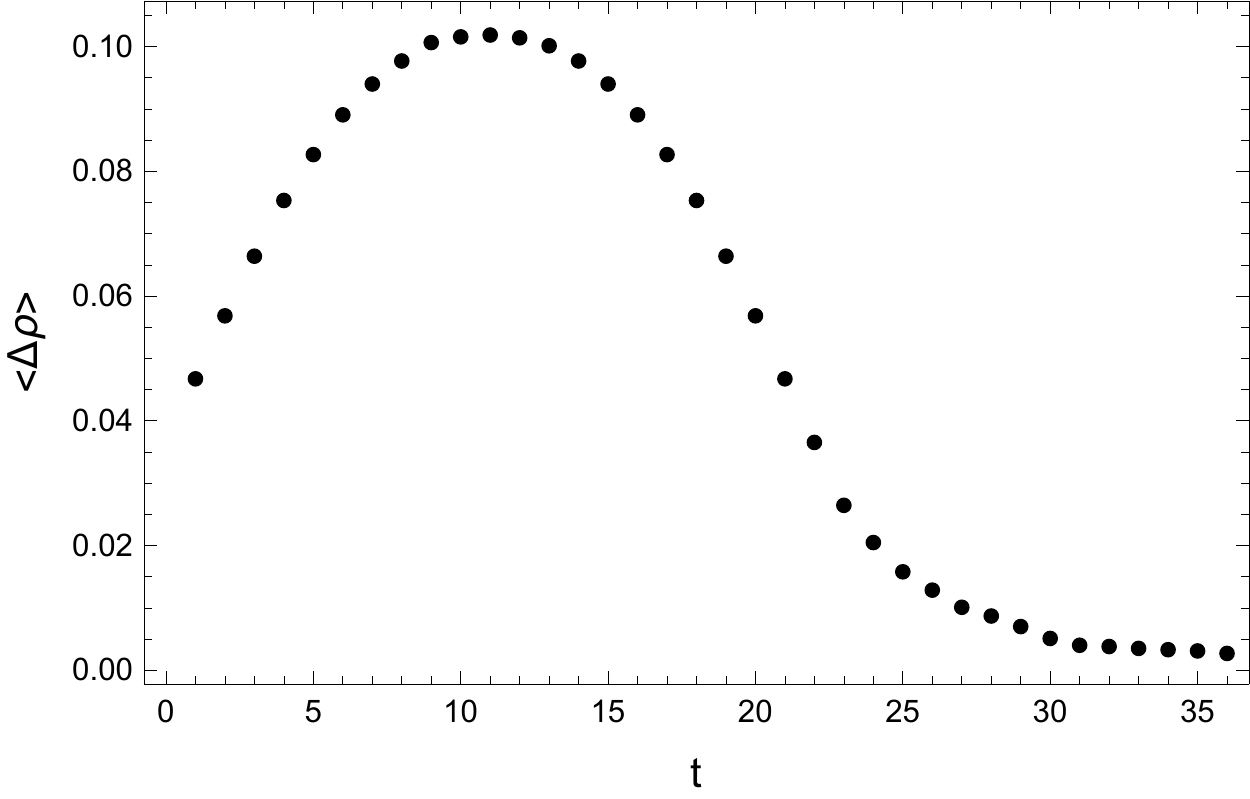}
\caption{\footnotesize The uncertainty of $\rho$ as a function of time $t$ that confirm how  the expectation  value $<\rho>_{t}$ is a genuine quantity.}
\label{fig:varianzarho}
\end{figure}

Let us consider now the full quantum evolution case (\ref{WdWe}). The absence of a potential term for the anisotropies suggests to us to consider for them a plane-waves solution, so that
\begin{equation}
\label{ondepiane}
\psi(\rho,\beta_{\pm}) = \frac{1}{2\pi}e^{ik_{+}\beta_{+}}e^{ik_{-}\beta_{-}}\chi(\rho),
\end{equation}
where $\{ k_{+},k_{-} \}$ are the quantum numbers associated to $\{\beta_{+},\beta_{-} \} $.
Taking into account this shape of the wave function in the Eq.(\ref{WdWe}) brings to the following differential equation:
\begin{equation}
\label{eqdiff1}
\left[\partial_{\rho}^{2} +\frac{k_{*}^2}{\rho ^{2}} -\Lambda_{*} \rho ^{2}\right]\chi( \rho) = E_{*} \chi(\rho),
\end{equation}
where
\begin{equation}
\label{costanti}
k_{*}^{2} = \frac{4}{9} (k_{+}^2 + k_{-}^2) \quad , \quad \Lambda_{*} = \frac{32 \pi^{2} \Lambda}{3} \quad , \quad E_{*} = \frac{32 \pi E}{3}. 
\end{equation}
Looking at Eq.(\ref{eqdiff1}) we can observe a formal analogy with the problem of the 3-D harmonic oscillator, where the angular momentum $l$ is in correspondence with the continuous values $k_{*}^2=-l(l+1)$. 
Following the analogy, we choose a solution for $\chi(\rho)$ of the form\cite{zettili}:
\begin{equation}
\label{forma}
\chi(\rho)= e^{-\frac{\sqrt{\Lambda_{*}}\rho^2}{2}}\rho^{\frac{1}{2} + \frac{\sqrt{1-4k_{*}^{2}}}{2}}\xi(\rho).
\end{equation}
The motivation of this choice is due to the fact that the terms $e^{\frac{\sqrt{\Lambda_{*}}\rho^2}{2}}$ and $\rho^{\frac{1}{2} + \frac{\sqrt{1-4k_{*}^{2}}}{2}}$ represent respectively the solutions of Eq.(\ref{eqdiff1}) in the limit $\rho \rightarrow \infty$ and $\rho\rightarrow 0$. The solution (\ref{forma}) should takes into account these two limit behaviours.
We assume a finite power series expansion for the function $\xi(\rho)$ of the form:
\begin{equation}
\label{espansioneserie}
\xi(\rho) = \sum\limits_{k=0}^{k'} c_{k,k'}\rho^{k} \quad , \quad k,k' \in 2 \mathbb{N}.
\end{equation}
The reason is due to the fact that this is the only way to obtain a physical acceptable solutions. Indeed, if we take into account a solution $\sum\limits_{k=0}^{\infty} c_{k}\rho^{k}$ for the problem (\ref{eqdiff1}) we obtain a non-converging series and then a diverging solution. To obtain a finite solution, as it is done in Eq.(\ref{espansioneserie}), we must required the series to be truncated at a certain power associated to $k'$.
Including expansion (\ref{espansioneserie}) in Eq.(\ref{eqdiff1}) we arrive to the following difference equation
\begin{multline}
c_{k+2,k'}(k+2)\left[ \sqrt{1-4 k_{*}^{2}} + k +2 \right] - \\ - c_{k,k'}\left[ E_{*} + \sqrt{\Lambda_{*}} \left( \sqrt{1-4 k_{*}^{2}} + 2k +2 \right)\right]=0.
\end{multline}
In order to obtain a finite solution we must set $c_{k+2,k'}=0$. This restriction allows us to determine the behaviour of the eigenvalue $E$, making use of the definitions (\ref{costanti}):
\begin{multline}
\label{autovalore}
 E_{*} + \sqrt{\Lambda_{*}} \left( \sqrt{1-4 k_{*}^{2}} + 2k' +2 \right) = 0 \Longrightarrow \\ \Longrightarrow E_{k',k_{\pm}} = -\frac{1}{4}\sqrt{\frac{3\Lambda}{2}}\left[ \sqrt{1-\frac{16}{9}(k_{+}^2 + k_{-}^2)} + 2k' +2 \right].
\end{multline}
In order to deal with a real eigenvalues, we consider a restriction for the values of $\{ k_{+},k_{-} \}$ of the form
\begin{equation}
(k_{+}^2 + k_{-}^2) \leq \frac{9}{16}.
\end{equation}
This way we obtain a spectrum for the eigenvalues that assumes only negative real values and then the associated dust-energy density is always positive.
Finally, always following the analogy with the 3-D harmonic oscillator, we can evaluate the coefficients $c_{k,k'}$ in terms of the $\Gamma$-function:
\begin{widetext}
\begin{equation}
c^{s}_{k,k'} = \frac{(-1)^{\frac{k}{2}} ( (-1)^{k} + 1) \Gamma \left[ 1+\frac{1}{2}\sqrt{1-\frac{16}{9}(k_{+}^2 + k_{-}^2)} \right] \left(\frac{32 \pi^{2} \Lambda}{3} \right)^{\frac{k}{4}}\frac{k'}{2}! }{ \Gamma \left[ 1+\frac{k}{2}\right] \Gamma \left[ 1+\frac{n}{2}+\frac{1}{2}\sqrt{1-\frac{16}{9}(k_{+}^2 + k_{-}^2)} \right] \left(\frac{k'}{2} - \frac{k}{2} \right)!}.
\end{equation}
\end{widetext}
Now we can obtain the shape of the entire wave function, solution to the problem (\ref{WdWe}), that is
\begin{equation}
\label{funzione d'onda}
\psi(\rho,\beta_{\pm}) = A e^{ik_{+}\beta_{+}}e^{ik_{-}\beta_{-}}e^{-\frac{\sqrt{\Lambda_{*}}\rho^2}{2}}\rho^{\frac{1}{2} + \frac{\sqrt{1-4k_{*}^{2}}}{2}}\sum\limits_{k=0}^{k'} c^{s}_{k,k'}\rho^{k},
\end{equation}
where $A$ is a normalization constant.
Now we want to perform a comparison between the classical trajectories (\ref{eq.moto rho}),(\ref{eq.moto beta}) and the expectation values of the associated operator $\hat{\rho}$ and $\hat{\beta_{\pm}}$. The states on which we evaluate them can be constructed taking into account the wave packets associated to the wave function (\ref{funzione d'onda}) peaked around classical values $\{ k^{'*},k_{+}^{*},k_{-}^{*} \}$,\textit{i.e.}
\begin{multline}
\label{pacc}
\Psi_{k^{'*},k_{\pm}^{*}}(\rho,\beta_{\pm}) = A\int\int_{R}dk_{\pm}e^{-\frac{(k_{+}-k_{+}^{*})^{2}}{2\sigma_{+}^{2}}}e^{-\frac{(k_{-}-k_{-}^{*})^{2}}{2\sigma_{-}^{2}}}\times \\ \times \sum\limits_{k'=1}^{\infty}  e^{-\frac{(k'-k^{'*})^{2}}{2\sigma ^{2}}}e^{-iE_{k',k_{\pm}}t}\psi(\rho,\beta_{\pm}),
\end{multline}
where the integrations on $\{ k_{+},k_{-} \}$ are restricted over the region $R=\{ k_{+},k_{-} \in \mathbb{R}| (k_{+}^2 + k_{-}^2) \leq \frac{9}{16} \}$ and we choose Gaussian weights to peak the wave packets.
The evolution in time of the expectation value of the operator $\hat{\rho}$ is evaluated over such states:
\begin{equation}
\label{valoremediorho}
<\hat{\rho}>_{t} = \int_{0}^{\infty} d\rho \int_{-\infty}^{\infty}d\beta_{\pm}(\Psi_{k^{'*},k_{\pm}^{*}})^{*}\rho\Psi_{k^{'*},k_{\pm}^{*}}.
\end{equation}
As we can see in Fig.(\ref{fig:pacchettirho}) we have an overlap between the expectation value (black points) and classical trajectory (red continuous line) only for late time $t$.  When we approach $t=0$, the expectation value moves away from the classical trajectory and it does not exhibit a singular behaviour. As a consequence, we can argue that in the evolutionary quantum model the singularity is avoided and it is replaced by a bounce.
The validity of this argument is supported by the analysis of the uncertainty:
\begin{multline}
\label{varianzarho}
<\Delta \rho^2>_{t} = \\ = \int_{0}^{\infty} d\rho \int_{-\infty}^{\infty}d\beta_{\pm}(\Psi_{k^{'*},k_{\pm}^{*}})^{*}\rho^{2}\Psi_{k^{'*},k_{\pm}^{*}} - <\hat{\rho}>_{t}^{2},
\end{multline}
essentially for two reasons.
The first one is, as we can see in Fig.(\ref{fig:varianzarho}), that when we are near the singularity the uncertainty $<\Delta \rho^2>$ has a maximum value  but it remains always small compared to the expectation value and it does not diverge in correspondence of the singularity. Thus we can conclude that the expectation value (\ref{valoremediorho}) is a good indicator for the system next to the singularity.
The second reason is the late times behaviour. It is clear from Fig.(\ref{fig:varianzarho}) that as we get farther away from the singularity, the uncertainty becomes smaller and smaller and approaches zero in the region of the overlap between expectation value and classical trajectory, guaranteeing that the Universe becomes always more and more classical at late times.
\begin{figure}[h!]
\centering
\includegraphics[scale=.67]{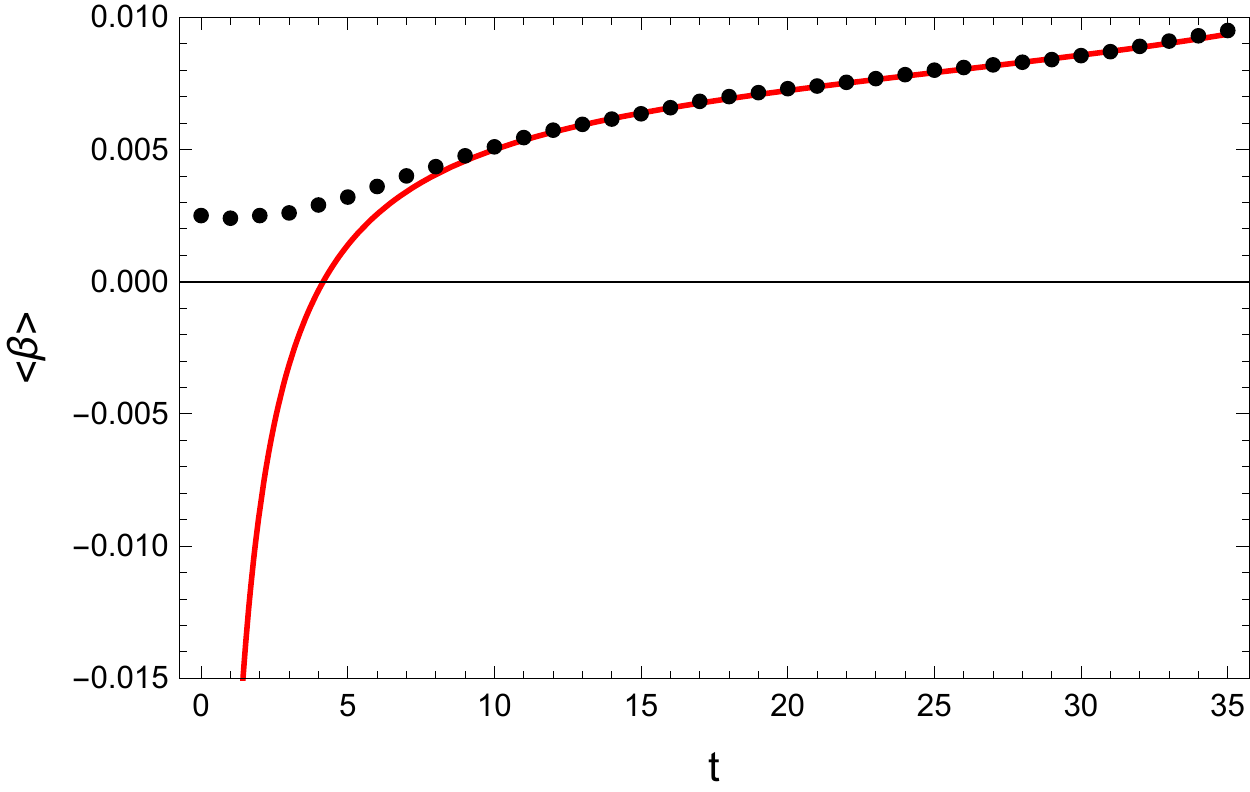}
\caption{\footnotesize  The black points represent the expectation value $<\beta_{\pm}>_{t}$ evaluated via numerical integration for the following choose of the integration parameters: $\Lambda =0.01, k^{'*}=5, k^{*}_{+}=k^{*}_{-}=0.1, \sigma_{+}=\sigma_{-}=0.01, \sigma=0.88$. The continuous red line represents the classical trajectory evaluated with the same classical parameters.}
\label{fig:pacchettibeta}
\end{figure}
\begin{figure}[h!]
\centering
\includegraphics[scale=.67]{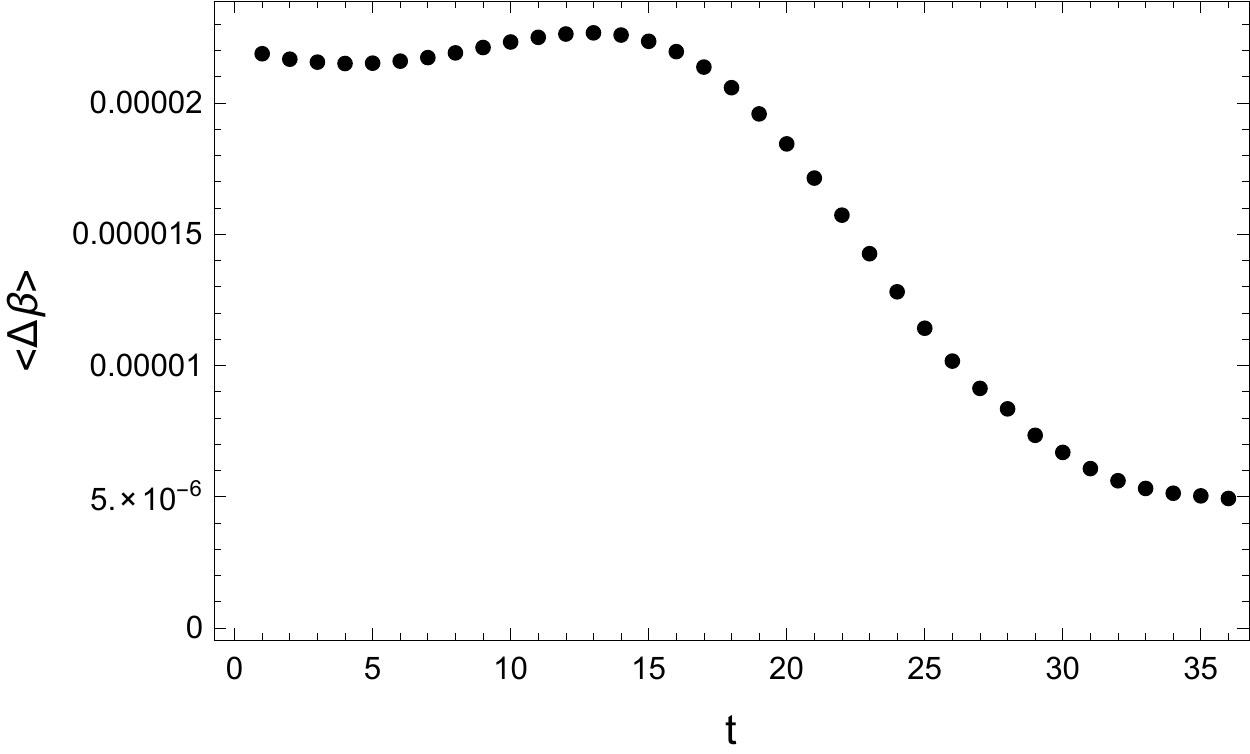}
\caption{\footnotesize The uncertainty of $\beta$ as a function of time $t$ that confirm how  the expectation  value $<\beta>_{t}$ is a genuine quantity.}
\label{fig:varianzabeta}
\end{figure}

The same approach can be used to compare expectation values related to the anisotropies with the corresponding classical trajectories. The evolution in time is:
\begin{equation}
\label{valoremediobeta}
<\hat{\beta_{\pm}}>_{t} =  \int_{0}^{\infty} d\rho \int_{-\infty}^{\infty}d\beta_{\pm}(\Psi_{k^{'*},k_{\pm}^{*}})^{*}\beta\Psi_{k^{'*},k_{\pm}^{*}}
\end{equation}

As we can see in Fig.(\ref{fig:pacchettibeta}) again we have an overlap between the expectation value (black points) and the classical trajectory (red continuous line) only for late time $t$. At early times, the diverging behaviour exhibited by the anisotropies at the classical level disappears in the quantum model. Indeed, when we approach the limit $t\rightarrow0$ the anisotropies remain small and finite. 
As before, the validity of this argument is supported by the analysis of the uncertainty $\Delta \beta_{\pm}$, defined as 
\begin{multline}
\label{varianzabeta}
<\Delta \beta^2>_{t} = \\ = \int_{0}^{\infty} d\rho \int_{-\infty}^{\infty}d\beta_{\pm}(\Psi_{k^{'*},k_{\pm}^{*}})^{*}\beta^{2}\Psi_{k^{'*},k_{\pm}^{*}} - <\hat{\beta}>_{t}^{2}.
\end{multline}
As it is shown in  Fig.(\ref{fig:varianzabeta}), the situation is exactly the same with respect to the case of the variable $\rho$, and this bring us to conclude in an analogous way that the (\ref{valoremediobeta}) is a genuine quantity to describe the system next to the singularity and to recover the classical limit for late times.
We conclude this section by noting how all the considerations here discussed for the initial singularity must remain valid when we consider the Bianchi I singularity in the future. By other words also the existing Big-Crounch is removed in favour of a bounce and our model acquires a cyclical feature. The non-diverging character of the anisotropies in this scenario can have intriguing implications for the so-called Big-Bounce cosmologies \cite{brandenberger}
in view of the possibility to minimize the effect on anisotropic evolution.

\section{Implication on the Bianchi IX model}
\label{sec:BIX}
Now, in order to implement the proprieties founded before to a general one model, we
analyse the Bianchi IX cosmology in presence of a negative cosmological constant in the context
of the evolutionary model. With respect to the configurational variables $ \{\rho, \beta_{+}, \beta_{-} \}$ the
superHamiltonian constraint takes the form
\begin{equation}
\label{hamrhoIX}
\mathcal{H} = -\frac{3}{32 \pi} p_{\rho}^{2} + \frac{p_{+}^{2} +p_{-}^{2}}{24 \pi \rho ^{2}}+ \frac{\pi}{2}\rho^{2/3}V_{IX}(\beta_{\pm}) -\pi \rho ^{2} \Lambda = E,
\end{equation}
where the potential term, which accounts for the spatial curvature of the model, reads as
\begin{multline}
\label{potanis}
 V_{IX}(\beta _{\pm})= e^{-8\beta _{+}}-4e^{-2\beta _{+}}\cosh (2\sqrt{3}\beta _{-})+ \\ +2e^{4\beta _{+}}\left[ \cosh(4\sqrt{3}\beta_{-})-1 \right].
\end{multline}
This potential is obtained selecting the three constants of structure $(\lambda_{l},\lambda_{m},\lambda_{n}) =(1,1,1)$ in the general potential expression in the Eq.(\ref{potenzialegenerale}).
As it is well known, in the context of the Misner-like variables, it is clear that the difference
between the Bianchi I model and the Bianchi IX model is the presence of the potential term $ \frac{\pi}{2}\rho^{2/3}V_{IX}(\beta_{\pm})$. For this reason we want to see if exists a regime in which the potential term of the Bianchi IX model is negligible with respect to the kinetic plus the cosmological constant term. In other words, we want to see when it is possible to argue that the properties founded in the previous section for the Bianchi I model are valid also for the Bianchi IX model. The importance to find a regime of this kind is due to presence of the BKL conjecture, which it
allows to extend such conclusion to the generic cosmological solution.
To this aim, we now want to assess the importance of the potential term $V^{*}_{IX}=	\frac{\pi}{2}\rho^{2/3}V_{IX}(\beta_{\pm})$ evaluated at the bounce as the dust energy $E$, estimated in the (\ref{autovalore}), changes.
\begin{figure}[h!]
\centering
\includegraphics[scale=.67]{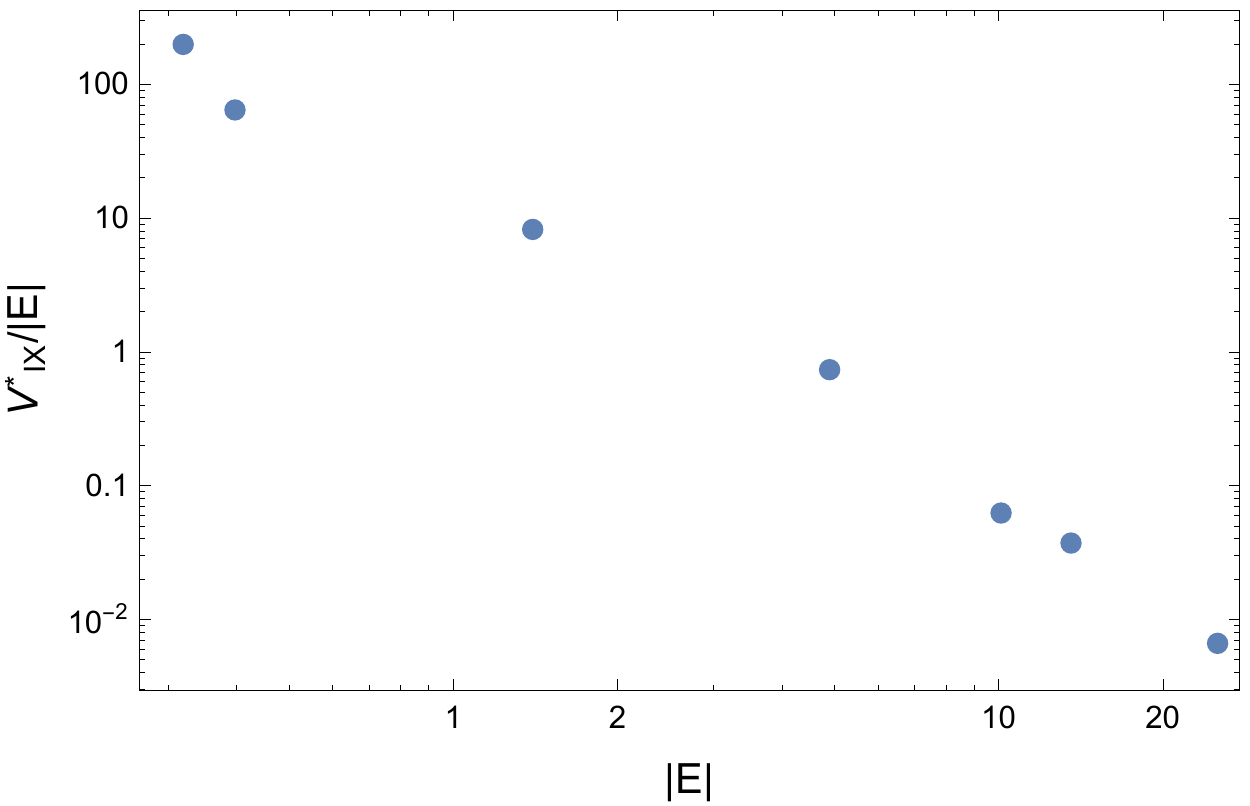}
\caption{\footnotesize The behaviour of the quantity $V_{IX}^{*}/|E|$ as a function of $|E|$ evaluated in correspondence of the bounce. The role of the Bianchi IX potential term became more and more marginal with the increase of the dust-energy.}
\label{fig:ve}
\end{figure}
As we can see in Fig.(\ref{fig:ve}), the potential term of the Bianchi IX model becomes more and more negligible as the magnitude of the dust energy density increases. This means that, following the trajectory of a Bianchi IX cosmology the relevant contribution comes from the kinetic plus cosmological constant term because the potential is more and more negligible as far as the parameter $E$ becomes large.
In this sense we can conclude, provided that the dust energy density is large enough to neglect the potential term, that the Bianchi IX model in presence of a negative cosmological constant in the evolutionary context possesses the same qualitatively features of the Bianchi I model previously founded.

\section{Phenomenological considerations}
\label{sec:phen}
Let us now provide a proper cosmological interpretation 
to the results we obtained in the previous sections and 
to outline the main merits of the proposed scenario. 

We considered a cosmological model which corresponds 
to the type I of the Bianchi classification, i.e. 
having zero spatial curvature and we included in the 
dynamics a small negative cosmological constant. 
The quantization of the model, to account for its 
behaviour nearby the cosmological singularity, has been 
performed accordingly to the Kucha\v{r}-Torre approach, relaying 
on a basic dualism between an external clock and the 
presence of a real dust fluid in the model evolution. 
The weak point of such a viable perspective to construct 
a physical time in quantum gravity, consists, in general, of the 
non positive definite nature of the dust energy 
density, emerging from the implementation of an 
external time (this fact reflects the non positive 
character of the super-Hamiltonian eigenvalue). 
However, in the considered model, this shortcoming 
of the dualism time-dust is fully overcome, since 
the energy of the dust is always positive and 
this is a consequence of the negative value of 
the cosmological constant, which, from a purely 
formal point of view, allows to compare the Universe 
volume quantum dynamics to an harmonic oscillator, 
but having a global minus sign. 

Then, studying the behaviour of quantum expectation values 
and uncertainties, we get the very surprising and 
valuable issue of a Big-Bounce cosmology. 
What makes our model physically meaning is the existence of 
a spontaneous classical limit, associated to the same 
harmonic structure removing the singularity. The quadratic 
potential, associated to the negative cosmological 
constant is responsible for a localization of the physical 
quantum states nearby the classical trajectory, as the 
Universe has a sufficiently large volume. 

This two important features of the model, i.e. the 
presence of a Big-Bounce nearby the classical location 
of the singularity and the natural classical limit 
of the expanded Universe, together with the turning point 
in the Universe late time evolution that the negative cosmological 
constant induces in the classical dynamics, suggests 
that our Bianchi I cosmology is an intriguing candidate 
for a cyclic Universe. 

This issue would be in itself a remarkable scenario, 
but our interest for the constructed picture is actually 
for the potential degree of generality it could represent. 
In fact, in section \ref{sec:BIX}, we have inferred that the 
behaviour of the Bianchi type I model can be extended, under 
suitable conditions (i.e. a sufficiently large value 
of the parameter $E$) to the most general Bianchi type 
IX cosmology, which is a good prototype for the generic 
cosmological Universe.  
By other words, it is a natural guess that the implementation of an evolutionary quantum gravity in the presence of a small 
negative cosmological constant can lead to a non-singular 
cyclic Universe even when we are referring it to a 
generic inhomogeneous Universe. According to the 
BKL conjecture \cite{BKL} and to its quantum implementation 
(the so-called long-wavelength assumption), for each sufficiently 
small neighbour of a space point, physically 
corresponding to the cosmological horizon, 
the dynamical evolution is qualitatively that one 
of a Bianchi IX Universe. Thus, we trace in the present 
analysis the basic dynamical features that could 
regularize the cosmological problem, without 
explicitly including an ultraviolet cut-off in the 
canonical Wheeler-DeWitt quantization of the system. 
Now, we should get light on the physical mechanism at the 
ground of such dynamical picture traced above and, 
in this respect, we investigate which of our ingredients 
is related in the model to a cut-off physics. 
\section{Physical interpretation of the Big Bounce}
\label{sec:poly}
In this section we want to show how is central the presence of the negative cosmological constant for the appearance of the Big Bounce. To this aim we analyse here an evolutionary Bianchi I model without the negative cosmological constant and we consider a cut-off polymer dynamics that makes discrete the isotropic variable $\rho$ in order to show how the behaviour of the quantum expectation values of the previous section and the behaviour of the polymer semiclassical dynamics are equivalent. This equivalence testifies  the fact that the negative cosmological constant plays the role of a cut-off physics. The model will be analysed in the same configurational space variables $\{ \rho,\beta_{+},\beta_{-}\}$ and the physical choice is to define the isotropic variable $\rho$ as a discrete variable and to leave unchanged the anisotropies $\{ \beta_{+},\beta_{-} \}$. We consider the polymer modification at a semiclassical level. It means that we are working with a modified superHamiltonian constraint obtained as the lowest order term of a WKB expansion for $\hbar\rightarrow 0$ of the full polymer quantum problem\cite{corichi},\cite{corichidue}.
This procedure formally consists in the replacement
\begin{equation}
\label{subpoly}
p_{\rho}^{2} \rightarrow \frac{2}{\mu ^{2}}[ 1 - \cos(\mu p_{\rho})],
\end{equation}
where $\mu$ is the polymer scale, or equivalently the lattice spacing for the variable $\rho$.
From the substitution (\ref{subpoly}) the superHamiltonian becomes
\begin{equation}
\label{SHp}
\mathcal{H}_{p} = -\frac{3}{16 \pi \mu^{2}}[ 1 - \cos(\mu p_{\rho})]+ \frac{p_{+}^{2} +p_{-}^{2}}{24 \pi \rho ^{2}},
\end{equation}
and again the superHamiltonian constraint is
\begin{equation}
\label{vincolopoly}
\mathcal{H}_{p} = E.
\end{equation}
As in the previous case, we can solve the semiclassical polymer dynamics making use of the Hamiltonian equations 
\begin{equation}
\label{eqhamiltonrho}
\begin{cases}& \dot{\rho}=\frac{\partial \mathcal{H}_{p}}{\partial p_{\rho}} = -\frac{3}{16\pi \mu}\sin(\mu p_{\rho}) \\ & \dot{p_{\rho}}=-\frac{\partial \mathcal{H}_{p}}{\partial \rho} = \frac{p_{+}^{2} +p_{-}^{2}}{12 \pi \rho ^{3}} \end{cases} 
\end{equation}
and of the constraint (\ref{vincolopoly}).
This way we obtain the following second order differential equation:
\begin{equation}
\label{eqdiffpolyrho}
\ddot{\rho}+\frac{(p_{+}^2+p_{-}^2) \left(1-\frac{2 \mu^2 (p_{+}^2+p_{-}^2)}{9 \rho^2}+\frac{16}{3} \pi  \mu^2 E \right)}{64 \pi ^2 \rho^3}=0
\end{equation}
It is not possible to individuate an analytical solution for the differential equation above, and then we perform a numerical integration. In order to find a link between the presence of a negative cosmological constant and the polymer scale we make a comparison between the classical and the quantum models analysed in Section \ref{sec:BI} and  this new semiclassical polymer model. We impose that the initial condition for the numerical integration of the differential equation (\ref{eqdiffpolyrho}) is exactly equal to $<\rho>_{0}$ adopted in Fig.(\ref{fig:pacchettirho}), \textit{i.e} we are arguing that the initial condition for the semiclassical evolutionary polymer problem matches the expectation value of the quantum evolutionary model in the correspondence of the bounce determined in the previous section. In order to perform this comparison we obviously choose the same classical values for the parameters $\{ p_{+},p_{-},E \}$ and the same corresponding parameters $\{ k^{'*},k_{+}^{*},k_{-}^{*} \}$ around which we have built the wave packets that we have used in Section \ref{sec:BI}. The only free parameter that we can fix is the polymer scale $\mu$.

\begin{figure}[h!]
\centering
\includegraphics[scale=.67]{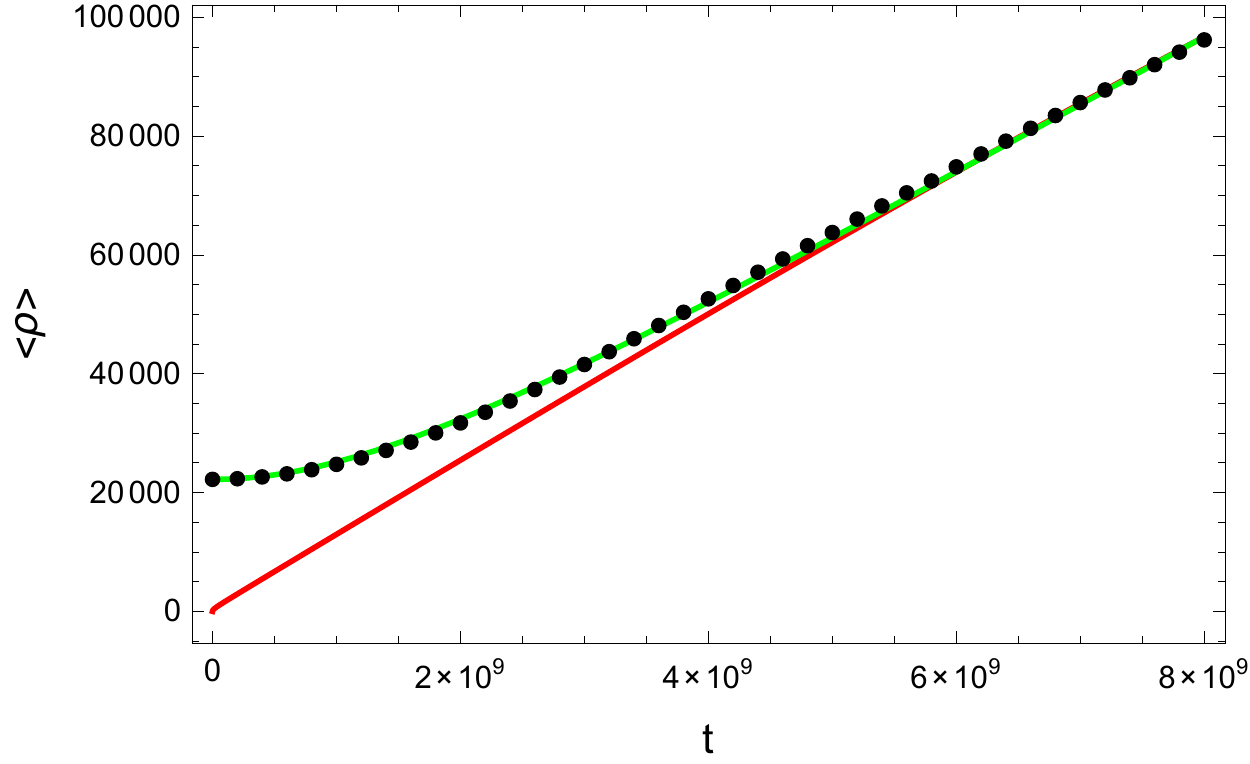}
\caption{\footnotesize The black points represent the expectation value $<\rho>_{t}$ evaluated via numerical integration for the following choose of the integration parameters: $\Lambda =10^{-20}, k^{'*}=20, k^{*}_{+}=k^{*}_{-}=0.1, \sigma_{+}=\sigma_{-}=0.01, \sigma=0.88$. The continuous red line represents the classical trajectory while the green line represents the semiclassical polymer trajectory, where the polymer scale is fixed with the choice $\mu=3.08\cdot 10^{5}$. }
\label{fig:poly}
\end{figure}

As we can see in Fig.(\ref{fig:poly}) it is possible to individuate a special value for the parameter $\mu$ for which the behaviour of $\rho(t)$ in the semiclassical polymer approach overlaps the expectation value $<\rho>_{t}$ in the quantum evolutionary theory. Furthemore, as it is expected for every kind of polymer approach, for late times the semiclassical polymer trajectory overlaps the classical one. This way we show that near the singularity in the context of the evolutionary theory, a negative cosmological constant acts the same way as a polymer modification related to the isotropic variable, \textit{i.e.} a cut-off physics.

\begin{figure}[h!]
\centering
\includegraphics[scale=.95]{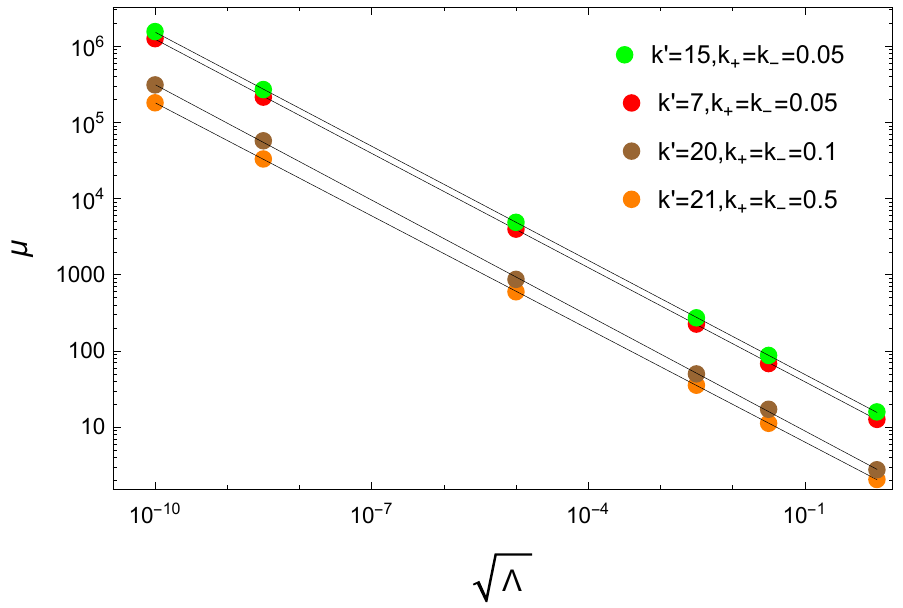}
\caption{\footnotesize The behaviour of the polymer scale $\mu$ as a function of $\log\sqrt{\Lambda}$. It is evident the existence of a law between the polymer scale and the negative cosmological constant, independently from the choice of the parameters.}
\label{fig:slope}
\end{figure}

It is possible to deepen the connection between the negative cosmological constant and the polymer scale making use of several numerical integrations related to different choice of the parameters values and seeing, time after time, if there is a general law. In Fig.(\ref{fig:slope}) it is shown the behaviour of $\log\mu$ as a function of $\log\sqrt{\Lambda}$, where the values of the numerical integration parameters $\{ k',k_{+},k_{-}\}$ used for evaluating the expectation value (\ref{valoremediorho})(and obviously the correspondent polymer integration parameters $\{p_{+},p_{-},E\}$) are fixed for each line. As we can see, the slope of the lines is always the same, independently from the choice of the parameters, and it is equal to $-\frac{1}{2}$. It means that a universal law exists such that:
\begin{equation}
\log \mu = \log \alpha_{k} - \frac{1}{2}\log \sqrt{\Lambda} \longrightarrow \mu ^{2}\sqrt{\Lambda} = \alpha_{k} ^{2},
\end{equation}
where the constant $\alpha_{k}=\alpha_{k',k_{+},k_{-}}$ depends on the values assigned to the parameters.
\section{Concluding remarks}
\label{sec:conclusion}
The main merit of the present work is 
in demonstrating how a rather general 
scenario for a cyclical Universe can 
be recovered even within the metric 
canonical quantum approach, as far 
as a well-defined evolutionary theory 
is taken into account. 

The basic ingredient of our approach is the small negative cosmological 
constant, which is responsible for 
the classical turning point, but overall, 
it induces an harmonic oscillator 
morphology in the quantum universe volume 
dynamics. 
The Bianchi I cosmology we addressed 
here allows the simultaneous manifestation of significant properties, like the 
Big-Bounce, the existence of well-defined classical limit and the positive 
character of the dust energy density, 
playing the role of a clock. 
However, what makes the present issues 
of intriguing cosmological meaning 
is the possibility to extend this 
picture to the Bianchi IX Universe. 
In fact, this property suggests that 
the considered minisuperspace scheme 
can be generalized to the generic 
inhomogeneous cosmological problem. 
As far as we implement the 
long-wavelength approximation to the 
inhomogeneous quantum dynamics, 
we can factorize the Wheeler superspace 
into the local minisuperspaces, associated 
to space point neighbours.  
From a physical point of view, we can 
speak of causal regions evolving, 
independently of each other, according 
to the non-singular cyclic dynamics 
we traced above. 
The implementation of the present ideas 
to a generic inhomogeneous Universe, 
as well as, the characterization of the 
role played by the matter, especially 
the radiation component, during the classical evolution, constitutes the
natural perspective of the present analysis. 

\section{Appendix: Derivation of the SuperHamiltonian for the Bianchi I model in the presence of a negative cosmological constant}
\label{appendix}
In this section we provide a brief derivation of the SuperHamiltonian (\ref{hamiltoniana}) and we will study the Bianchi I and Bianchi IX model, respectively the simplest and most general homogenous but anisotropic model. A generic homogeneous model with space-time metric $g_{ij}$ has to preserve the invariance of the spatial line element under suitable group of transformations. It means that the spatial line element
\begin{equation}
dl^{2}=h_{\alpha \beta}(t,x)dx^{\alpha}dx^{\beta},
\end{equation}
under the isometry $T:x\rightarrow x'$, has to left invariant the 3-dimensional metric $h_{\alpha \beta}(t,x)$ so that in the transformed line element
\begin{equation}
dl^{2}=h_{\alpha \beta}(t,x')dx'^{\alpha}dx'^{\beta}
\end{equation}
the spatial metric $h_{\alpha \beta}(t,x') = h_{\alpha \beta}(t,x)$.
If we introduce three spatial vectors $ \{ l(x),m(x),n(x) \}$ that satisfy the homogeneity condition, the metric $h_{\alpha\beta}$ can be expressed in the form
\begin{equation}
h_{\alpha \beta} = a^{2}(t)l_{\alpha}l_{\beta} + b^{2}(t)m_{\alpha}m_{\beta} + c^{2}(t)n_{\alpha}n_{\beta},
\end{equation}
where $a(t), b(t), c(t)$ are three different
cosmic scale factors along the three spatial directions. Consequently, the vacuum Einstein equations in a synchronous reference and for a generic homogeneous cosmological model are
\begin{equation}
\label{einsteinequ}
\begin{cases}
& -R^{l}_{l} = \frac{(\dot{a}bc)^{.}}{abc} + \frac{1}{2(abc)^{2}}\left[ \lambda _{l}^{2}a^{4} - (\lambda _{m}b^{2} - \lambda _{n}c^{2})^{2} \right] = 0 \\
& -R^{m}_{m} = \frac{(a\dot{b} c)^{.}}{abc} + \frac{1}{2(abc)^{2}}\left[ \lambda _{m}^{2}b^{4} - (\lambda _{l}a^{2} - \lambda _{n}c^{2})^{2} \right] = 0 \\
& -R^{n}_{n} = \frac{(ab\dot{c} )^{.}}{abc} + \frac{1}{2(abc)^{2}}\left[ \lambda _{n}^{2}c^{4} - (\lambda _{l}a^{2} - \lambda _{m}b^{2})^{2} \right] = 0 \\
& -R^{0}_{0} = \frac{\ddot{a}}{a}+ \frac{\ddot{b}}{b} +\frac{\ddot{c}}{c} = 0.
\end{cases}
\end{equation}
The constants  $(\lambda_{l},\lambda_{m},\lambda_{n})$ are called \textit{constants of structure} and they can only assume the values $(-1,0,1)$. The form of Eq.'s (\ref{einsteinequ}) takes into account the dynamics of the homogeneous models that are relevant near the singularity. In particular we can only consider, in the Eq.'s (\ref{einsteinequ}), the behaviour of six models, called Bianchi I, II, VI, VII, VIII, IX, that belong to the so-called \textit{Bianchi Classification}\cite{landau}. This classification contains the all nine possible models that respect the homogeneity constraint in the same way as $K =\{-1,0,1\}$ identifies the possible symmetry types for homogeneous and isotropic FRW three-spaces. In particular three of them, the Bianchi I, the Bianchi V and the Bianchi IX model represent the anisotropic generalization of the flat, open and closed FRW metrics respectively.

Let us consider now a line element for a generic homogeneous space-time in the ADM (Arnowitt-Deser-Misner \cite{ADM}) form:
\begin{equation}
\label{metricaADM}
ds^{2} = N(t)^{2}dt^{2} - h_{\alpha\beta}dx^{\alpha}dx^{\beta},
\end{equation}
where $N(t)$ is the lapse function and where we redefined the three scale factors $\{a(t),b(t),c(t)\}$ in such a way to have a spatial line element of the form:
\begin{multline}
dl^{2}=h_{\alpha \beta}dx^{\alpha}dx^{\beta} = \\ =( e^{q_{l}}l_{\alpha}l_{\beta} + e^{q_{m}}m_{\alpha}m_{\beta} + e^{q_{n}}n_{\alpha}n_{\beta}) dx^{\alpha}dx^{\beta} =\eta_{ab}\omega^{a}\omega^{b},
\end{multline}
where we introduced the matrix $\eta_{ab}=diag\{e^{q_{l}},e^{q_{m}},e^{q_{n}} \} $ and a set of three invariance form $\omega^{a}=\omega^{a}_{\alpha}dx^{\alpha}$ with $\omega^{a}_{\alpha} = \{ l_{\alpha},m_{\alpha},n_{\alpha} \}$.

In order to introduce the dynamical character of the gravitational field let us consider the Einstein-Hilbert Action in the presence of a negative cosmological constant:
\begin{equation}
S = -\frac{1}{2\kappa}\int d^{4}x\sqrt{-g}(R-2\Lambda),
\end{equation}
where\footnote{For the calculation in this Appendix we use the natural units $c=\hbar=1$} $\kappa=8\pi G$ and $R$	is the \textit{Ricci scalar}.
Let us start by studying the variation of the previous action. It means that we have to evaluate the determinant of the space time metric and the Ricci Scalar for the particular case of the homogeneous space-time represented in the line element \ref{metricaADM}. When we do this we obtain:
\begin{equation}
\label{azioneBIX}
\delta S_{g} = \delta \int_{t_{1}}^{t_{2}}\mathcal{L}(q_{a},\dot{q_{a}})dt = 0
\end{equation}
where $t_{1}$ and $t_{2}$ are two fixed instants of time with $t_{1}<t_{2}$ and the Lagrangian $\mathcal{L}$ is of the form
\begin{equation}
\label{lagBIX}
\mathcal{L} = -\frac{8\pi ^{2}\sqrt{\eta}}{\kappa}\left[\frac{1}{2N}(\dot{q}_{l}\dot{q}_{m} + \dot{q}_{l}\dot{q}_{n} + \dot{q}_{m}\dot{q}_{n}) - N \overline{R} + N\Lambda \right].
\end{equation}
In the Lagrangian (\ref{lagBIX} we introduce the quantity $\eta = det(\eta_{ab})=e^{q_{l}+q_{m}+q_{n}}=e^{\sum _{a}q_{a}}$ while $\overline{R}$ represents the 3-dimensional Ricci Scalar and it is connected with the constants of structure in such a way that
\begin{equation}
\label{pot}
\eta \overline{R} = -\frac{1}{2}\left[ \sum _{a}\lambda ^{2} _{a}e^{2q_{a}}-\sum _{a\neq b}\lambda _{a}\lambda _{b}e^{q_{a}+q_{b}} \right],
\end{equation}
where the indexes $\{a,b\}$ take values in $\{l,m,n\}$.
The choice of the constants of structure that appear in the  Eq.(\ref{pot}) determines the particular homogeneous model that we can select inside the Bianchi Classification.

From the Lagrangian (\ref{lagBIX}) we can obtain the Hamiltonian of the system performing a Legendre transformation. The conjugated momentas to the generalized coordinate $q_{a}$ are the following
\begin{equation}
\label{momq}
\begin{cases}
& p_{l}=\frac{\partial \mathcal{L}_{g}}{\partial \dot{q}_{l}} = -\frac{4\pi ^{2}\sqrt{\eta}}{kN}(\dot{q}_{m}+\dot{q}_{n}) \\
& p_{m}=\frac{\partial \mathcal{L}_{g}}{\partial \dot{q}_{m}} = -\frac{4\pi ^{2}\sqrt{\eta}}{kN}(\dot{q}_{n}+\dot{q}_{n})\\
& p_{n}=\frac{\partial \mathcal{L}_{g}}{\partial \dot{q}_{n}} = -\frac{4\pi ^{2}\sqrt{\eta}}{kN}(\dot{q}_{l}+\dot{q}_{m})
\end{cases}
\end{equation}
and taking into account the transformation 
\begin{equation}
N\mathcal{H} = \sum _{a=l,m,n}p_{a}\dot{q}_{a}-\mathcal{L}_{g},
\end{equation}
where $\mathcal{H}$ is the SuperHamiltonian of the system, we can put the action in the form
\begin{equation}
\label{azionecoogen}
S_{g} = \int dt \left(\sum _{a} p_{a}\dot{q}^{a} - N\mathcal{H}\right)
\end{equation}
with
\begin{equation}
\label{superH}
\mathcal{H}=\frac{k}{8\pi ^{2}\sqrt{\eta}}\left[\sum _{a}(p_{a})^{2}-\frac{1}{2}\left(\sum _{b}p_{b}\right)^{2}-\frac{64\pi ^{4}}{k^{2}} \left( \eta \overline{R} + \eta \Lambda \right) \right].
\end{equation}

A very useful set of generalized coordinates is represented by the Misner Variables $\{\alpha, \beta_{+}, \beta_{-}\}$\cite{misner}\cite{misnermixmaster},\textit{i.e.}
\begin{equation}
\label{varmisner}
\begin{cases}
q_{l} = 2(\alpha + \beta_{+}+\sqrt{3}\beta{-}) \\ 
q_{m} = 2(\alpha + \beta_{+}-\sqrt{3}\beta{-}) \\ 
q_{n} = 2(\alpha -2 \beta_{+}).
\end{cases}
\end{equation}
Respect to the Misner variables the metric $\eta_{ab}$ assumes the form
\begin{equation}
\label{eta misner}
\eta_{ab} = e^{2\alpha}(e^{2\beta})_{ab} \rightarrow det(\eta_{ab}) = e^{6\alpha}
\end{equation} 
It is possible to show that the variable $\alpha$ represents the isotropic component of the Universe, being related to the volume, while the matrix $\beta_{ab} = diag(\beta _{+} + \sqrt{3}\beta _{-} , \beta _{+} - \sqrt{3}\beta _{-} , -2\beta _{+})$ accounts for the anisotropy of the system. In terms of this new variables the action (\ref{azionecoogen}) takes the form
\begin{equation}
\label{azionecoomis}
S_{g} =\int \left( p_{\alpha}\dot{\alpha}+p_{+}\dot{\beta}_{+}+p_{-}\dot{\beta }_{-} - N\mathcal{H} \right)dt
\end{equation}
where
\begin{equation}
\label{Hmisner}
\mathcal{H}= \frac{k}{3(8\pi)^{2}}e^{-3\alpha}(-p_{\alpha}^{2}+p_{+}^{2}+p_{-}^{2}+\mathcal{V})-\frac{8 \pi^{2} \Lambda}{\kappa}e^{3\alpha}
\end{equation}
and the scalar curvature term becomes
\begin{equation}
\label{potmisner}
\mathcal{V}=-\frac{6(4\pi)^{4}}{k^{2}}\eta \overline{R} = \frac{3(4\pi)^{4}}{k^{2}}e^{4\alpha}V(\beta _{\pm}).
\end{equation}
The potential term $V(\beta_{\pm})$ accounts for spatial curvature of the model and is given by the expression:
\begin{multline}
\label{potenzialegenerale}
V(\beta_{\pm})=\lambda_{l}^{2}\left(e^{-8\beta_{+}-2e^{4\beta_{+}}}\right) + \\ +\lambda_{m}^{2}\left(e^{+4(\beta_{+}+\sqrt{3}\beta_{-})} - 2e^{-2(\beta_{+}-\sqrt{3}\beta_{-})}\right) + \\ +\lambda_{n}^{2}\left(e^{+4(\beta_{+}-\sqrt{3}\beta_{-})} - 2e^{-2(\beta_{+}+\sqrt{3}\beta_{-})}\right)
\end{multline}

When we choose the Bianchi I model we select an homogeneous model with the three constant of structure equal to zero, or in order words we are taking into account a model with zero spatial curvature. When we do this the Hamiltonian (\ref{Hmisner}) simply becomes
\begin{equation}
\label{finale}
\mathcal{H}_{I}= \frac{k}{3(8\pi)^{2}}e^{-3\alpha}(-p_{\alpha}^{2}+p_{+}^{2}+p_{-}^{2}) - \frac{8 \pi^{2} \Lambda}{\kappa}e^{3\alpha}.
\end{equation}
If now we make explicit $k=8 \pi G$ in the "Geometrical Unit", so $(c=G=\hbar=1)$, the SuperHamiltonian (\ref{finale}) reduce to the SuperHamiltonian in the Eq.(\ref{hamiltoniana}).

Finally, for $\lambda_{m}=\lambda_{m}=\lambda_{m}=1$ we get the Bianchi IX model and the potential (\ref{potmisner}).
 \section*{Acknowledgments}
 S.C. acknowledges the support of {\it Istituto Nazionale di Fisica Nucleare} IS-QGSKY and IS-TEONGRAV.

\end{document}